\documentclass{emulateapj}
\usepackage{apjfonts,times}

\newcommand{\beq}{\begin{equation}}
\newcommand{\eeq}{\end{equation}}

\def\alp{\mbox{$\alpha$}}

\def\farcm{\hbox{$.\mkern-4mu^\prime$}}
\def\farcs{\hbox{$.\!\!^{\prime\prime}$}}
\def\earth{\hbox{$\oplus$}}

\def\arcsec{\hbox{$^{\prime\prime}$}}
\def\solar{\mbox{$_{\normalsize\odot}$}}

\newcommand{\AmS}{{\protect\the\textfont2
  A\kern-.1667em\lower.5ex\hbox{M}\kern-.125emS}}
\newcommand{\lsim}{\ \raise
-2.truept\hbox{\rlap{\hbox{$\sim$}}\raise5.truept\hbox{$<$}\ }}
\newcommand{\gsim}{\ \raise
-2.truept\hbox{\rlap{\hbox{$\sim$}}\raise5.truept\hbox{$>$}\ }}
\newcommand{\simsim}{\ \raise
-2.truept\hbox{\rlap{\hbox{$\sim$}}\raise5.truept\hbox{$\sim$}\ }}

\slugcomment{Accepted for Publication in ApJ, May 1, 2007}

\shorttitle{Clustered Star Formation in NGC~602/N~90 in the SMC 
with Spitzer/IRAC Photometry}
\shortauthors{D. A. Gouliermis, S. P. Quanz, \& Th. Henning}

\begin{document}

\title{Clustered Star Formation in the Small Magellanic Cloud. A
{\em Spitzer}/IRAC View of the Star-Forming Region
NGC~602/N~90\altaffilmark{1}}


\author{Dimitrios A. Gouliermis, Sascha P. Quanz, \& Thomas Henning}
\affil{Max-Planck-Institute for Astronomy, K\"onigstuhl 17, 69117
Heidelberg, Germany}

\email{[dgoulier], [quanz], [henning]@mpia-hd.mpg.de}

\altaffiltext{1}{Research supported by the Deutsche
Forschungsgemeinschaft (German Research Foundation)}

\begin{abstract}

We present {\em Spitzer}/IRAC photometry on the star-forming {\sc H~ii}
region N~90, related to the young stellar association NGC~602 in the
Small Magellanic Cloud. Our photometry revealed bright mid-infrared
sources, which we classify with the use of a scheme based on templates
and models of red sources in the Milky Way, and criteria recently
developed from the {\em Spitzer} Survey of the SMC (Bolatto et al. 2007)
for the selection of candidate Young Stellar Objects (YSOs). We detected
57 sources in all four IRAC channels in a 6\farcm2 $\times$ 4\farcm8
field-of-view centered on N~90; 22 of these sources are classified as
candidate YSOs. We compare the locations of these objects with the
position of optical sources recently found in the same region with
high-resolution HST/ACS imaging of NGC~602 by Schmalzl et al. (2007),
and we find that 17 candidate YSOs have one or more optical
counterparts. All of these optical sources are identified as pre-main
sequence stars, indicating, thus, ongoing clustered star formation
events in the region. The positions of the detected YSOs and their
related PMS clusters give a clear picture of the current star formation
in N~90, according to which the young stellar association photo-ionizes
the surrounding interstellar medium, revealing the {\sc H~ii} nebula,
and triggering sequential star formation events mainly along the eastern
and southern rims of the formed cavity of the parental molecular cloud. 

\end{abstract}


\begin{figure*}[t!]
\vspace*{2.5truecm}
\begin{center}
{\large The postscript file is removed due to size limitations. \\
To see the figure download the corresponding JPG file: ``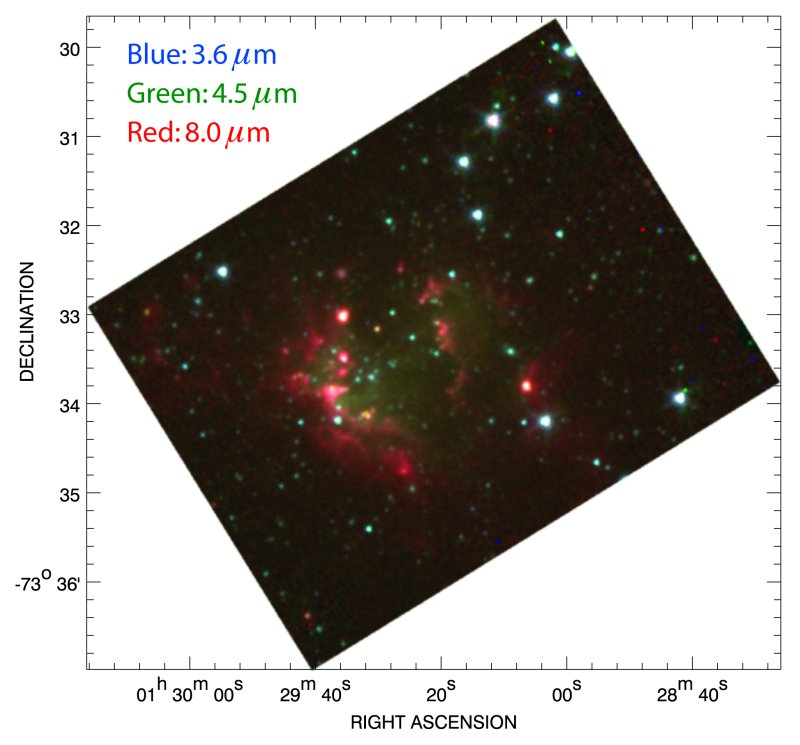''}
\end{center}
\vspace*{2.5truecm}
\caption{{\em Spitzer} image of the region surrounding NGC~602/N~90.
IRAC three-color image is shown, with IRAC-1 3.6 \micron\ in blue,
IRAC-2 4.5 \micron\ in green, and IRAC-4 8 \micron\ in red.}
\label{irac-maps}
\end{figure*}

\section{Introduction}

The {\em Spitzer} Space Telescope opens a unique mid- and far-infrared
window to the exploration of massive star formation not only in our
galaxy but also in our neighboring dwarf galaxies, the Large and Small
Magellanic Cloud (LMC, SMC). Stellar associations in both the Magellanic
Clouds (MCs) contain the richest sample of young bright stars in these
galaxies, and consequently our knowledge on their young massive stars
has been collected from studies of such stellar systems (see e.g. Massey
2002). Furthermore, almost every MCs association coincide with one or
more {\sc H~ii} regions, as they have been earlier cataloged (Henize
1956; Davies et al. 1976), and it has been shown that the measured
H\alp\ fluxes are in excellent agreement with those expected from the
ionizing flux of the detected massive stars (Massey 1993). However, MCs
associations are not mere aggregates of young bright stars alone, but
they also host large numbers of faint pre-main sequence (PMS) stars, as
recent HST studies showed for both the LMC (Gouliermis et al. 2006) and
the SMC (Nota et al. 2006). Both, the existence of {\sc H~ii} regions
and PMS stars in stellar associations of the MCs indicate that star
formation may be still active in their vicinity, and {\em Spitzer}
offers a unique opportunity to test this hypothesis.

Indeed, using {\em Spitzer} observations, Jones et al. (2005) reported
four Young Stellar Objects (YSOs) in the {\sc H~ii} complex N 159
(located in the vicinity of the association LH 105; Lucke \& Hodge 1970)
in the LMC. Chu et al. (2005) obtained {\em Spitzer} observations of the
super-bubble N 51D (associations LH~51/54) in the LMC and they found
three YSOs projected within the super-bubble. Images from the {\em
Hubble} Space Telescope (HST) in H\alp\ and [{\sc S ii}] allowed them to
identify dust globules associated with two of them. The third YSO was
associated with the first Herbig-Haro object detected outside the
Galaxy. Furthermore, Meixner et al. (2006) in their presentation of the
first results from the {\em Spitzer} Space Telescope Survey of the LMC
Legacy Project {\em Surveying the Agents of a Galaxy's Evolution}
(SAGE), show as a test-case epoch 1 data of the area of two {\sc H~ii}
regions, N 79/N 83 (associations LH 2/LH 5). They found that MIPS 70
\micron\ and 160 \micron\ images of the diffuse dust emission of the
regions reveal a distribution similar to the gas emissions, especially
the {\sc H i} 21 cm emission, and they note that the measured
point-source sensitivity for the epoch 1 data is consistent with
expectations for the survey.

Bolatto et al. (2007), recently, presented their results from the {\em
Spitzer} Survey of the Small Magellanic Cloud (S$^3$MC), which imaged
the star-forming body of the SMC in all seven MIPS and IRAC wave bands.
These authors compiled a photometric catalog of 400,000 mid- and
far-infrared point sources in the SMC, from which they identified 282
bright candidate YSOs as bright 5.8 \micron\ sources with very faint
optical counterparts and very red mid-infrared colors ([5.8]$-$[8.0] $>$
1.2 mag). More recently, Simon et al. (2007) used the same observations
and applied spectral energy distribution (SED) fits based on models by
Whitney et al. (2003a, b, 2004) to enhance the catalog of candidate YSOs
in the brightest {\sc H~ii} region in the SMC, N~66.

The present paper deals with mid-infrared imaging from the {\em Spitzer} 
Space Telescope of the {\sc H~ii} region LHA~115-N90 (Henize 1956) or in 
short N~90, located in the wing of the SMC. This region, also cataloged by 
Davies et al. (1976) and named DEM~S~166, is related to the young star 
cluster NGC~602, which was classified as a stellar association by Hodge 
(1985; system No 68 in his catalog). We present the results of our 
investigation on the candidate YSOs in the region of NGC~602/N~90, 
identified with our photometry according to criteria based on galactic 
templates and recent {\em Spitzer} observations of other star forming 
regions in the SMC.

Specifically, in \S 2 we discuss the retrieval of the {\em Spitzer} data 
used in our study and their photometry. General description of the 
mid-infrared characteristics of the region is given in \S 3, and in \S 4 
the point sources detected with our photometry, and the selection of 
candidate YSOs are presented. The recent release of high-resolution data 
on NGC~602/N~90 obtained with the {\em Advanced Camera for Surveys} on 
{\em Hubble} Space Telescope allowed us a detailed photometric study of 
the stellar association and its surroundings down to the faint red 
pre-main sequence stars of the system (Schmalzl et al. 2007). In \S 4 we 
also combine the results from both {\em Hubble} and {\em Spitzer} 
photometries to identify the optical counterparts of the candidate YSOs 
found in the region. Based on our findings we discuss the current star 
forming process that probably took place in the region of NGC~602/N~90 
with emphasis to the clustered behavior of star formation in \S 5. 
Finally, general conclusions of this study are given in \S 6.

\begin{figure*}[t!]
\epsscale{1.}
\plotone{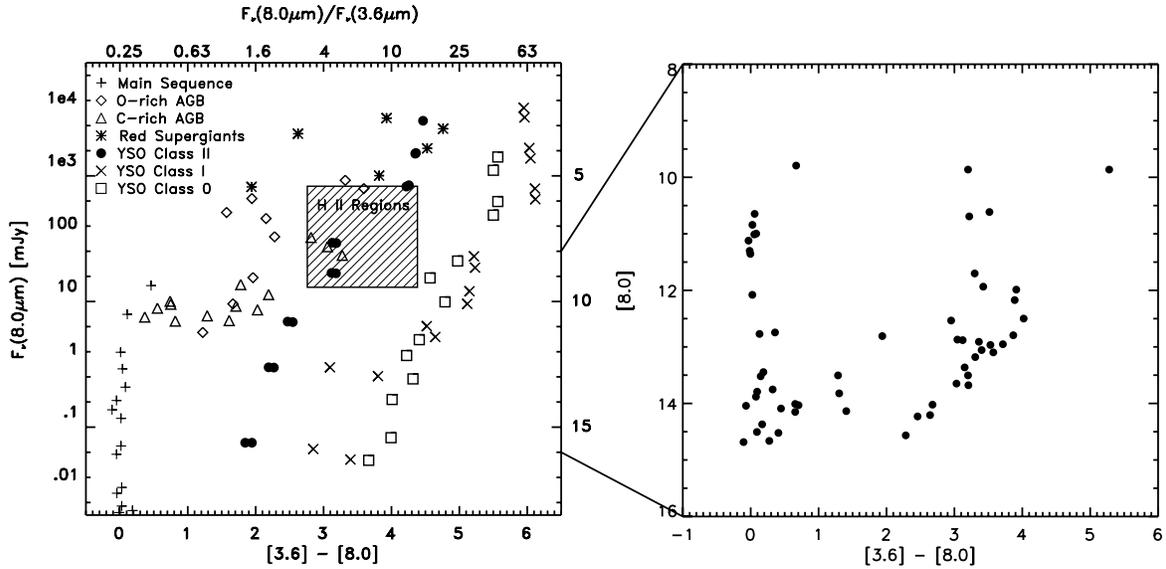}
\caption{{\em Left}: IRAC CMD of template/model photometry of Cohen (1993)
and Whitney et al. (2004) for IR sources in the Milky Way. Symbols code
different types of sources throughout the SMC: Main-sequence O stars,
RSGs, O-rich and C-rich AGB stars, {\sc H~ii} regions, and YSOs. {\em
Right}: The corresponding CMD of sources found in the region of
NGC~602/N~90 with our IRAC photometry. The comparison of these CMDs
denotes that there are 27 objects in our photometry (located at the right
part of the CMD on the right with colors ${\rm [3.6]}-{\rm [8.0]} >
2$~mag) that match the type of galactic YSOs.
\label{iractempl}}
\end{figure*}

\section{Data Retrieval and Photometry}

The imaging data of our study have been obtained with the {\em Infrared
Array Camera} (IRAC; Fazio et al. 2004) on-board {\em Spitzer} Space
Telescope (Werner et al. 2004). The images are publicly available and
are downloaded from the {\em Spitzer} Space Telescope Archive using the
{\tt Leopard} package\footnote{{\tt Leopard} can be downloaded from {\tt
http://ssc.spitzer.caltech.edu/propkit/spot/}.}. These observations
(dataset archive ID: AOR 12485120) were obtained within the GTO Science
Program 125 (PI: G. Fazio) on 28 November 2004. We retrieved the Post
Basic Calibrated Data ({\tt pbcd}) for all four IRAC channels (3.6
\micron, 4.5 \micron, 5.8 \micron\ and 8.0 \micron), which are already
fully co-added and calibrated. After visual inspection of the images we
carried out PSF-photometry with the {\tt daophot} package provided
within the {\tt IRAF\footnote{{\tt IRAF} is distributed by the National
Optical Astronomy Observatories, which are operated by the Association
of Universities for Research in Astronomy, Inc., under cooperative
agreement with the National Science Foundation. ({\tt
http://iraf.noao.edu/})}} environment. In the case of the region of
NGC~602/N~90, PSF photometry is required as the field is too crowded for
aperture photometry to be appropriately applied.  Consequently, for each
filter we constructed a reference PSF by combining the PSF of at least 5
objects with a high signal to noise ratio and we used this PSF to fit
all objects identified with a 3-$\sigma$ confidence level over the local
background.

We converted the pixel values in each filter from MJy\,sr$^{-1}$ to
DN\,s$^{-1}$ according to ``IRAC Data Handbook'' version 5.1.1 and we
computed the corresponding magnitudes as $m~=-2.5\,\log{(f)}+\Delta_{\rm
ZP}$ with $f$ denoting the flux measured in DN\,s$^{-1}$ and
$\Delta_{\rm ZP}$ being the zero point for each filter. Zero points were
taken from Hartmann et al. (2005): 19.66 (3.6 \micron), 18.94 (4.5
\micron), 16.88 (5.8 \micron) and 17.39 (8.0 \micron). For the initial
PSF fit we used a PSF size of 2 pixel, and applied aperture corrections
as described in the ``IRAC Data Handbook'' to obtain the final
magnitudes. Finally, we cross-matched the obtained catalogs for each
filter to identify the sources that are detected in more than one IRAC
bands, by allowing a maximum offset of 2 pixels for the center of each
point source.

\section{Description of the Mid-Infrared Observations of NGC~602/N~90}

{\em Spitzer} covers a suitable wavelength range (3.6 - 160 \micron) for 
defining the dust properties of the interstellar medium (ISM). The 
mid-infrared channels of IRAC (3.6, 4.5, 5.8 and 8 \micron) are good 
tracers of the smallest grains through Polycyclic Aromatic Hydrocarbon 
(PAH) emission. Very small grains (typical sizes 12 - 150 \AA) and PAHs 
(typical sizes 4 - 12 \AA; Desert et al. 1990) are very efficient in 
heating the gas through the photoelectric effect (Bakes \& Tielens 1994) 
and therefore their analysis is very important for understanding the 
thermodynamics of the ISM.

IRAS data on both LMC and SMC suggest a deficit of the smallest dust 
grains in the low-metallicity environments of these galaxies due to the 
lower 12 \micron\ emission in comparison to the Milky Way and other higher 
metallicity galaxies (Schwering 1989; Sauvage et al. 1990). A study in the 
SMC by Stanimirovi\'{c} et al. (2000) concluded that the dust in the SMC 
is dominated by large grains. The lower abundance of PAHs and very small 
grains in the SMC is also supported by detailed modeling of SMC UV 
absorption and FIR emission data by Weingartner \& Draine (2001), Li \& 
Draine (2002), and Clayton et al. (2003). A different grain size 
distribution with respect to the Galaxy, or grain destruction, possibly 
due to the intense UV radiation and/or by supernova shocks have been 
proposed to explain the very small grain paucity in both galaxies. The 
absence of PAHs and very small grains has a profound influence on the gas 
heating and the existence of cold and warm phases in the interstellar 
medium (Wolfire et al. 1995).

The images of NGC~602/N~90 taken with {\em Spitzer}, which we present
here, can be used for a qualitative analysis on the dust properties and
its relation to stellar sources of radiation in the region of NGC~602. A
color composite image from IRAC channels 1, 2 and 4 is shown in Fig. 
\ref{irac-maps}.  The 3.6 and 4.5 \micron\ wave-bands are mostly
sensitive to stellar photospheres and very hot circumstellar dust, and
therefore the blue (3.6 \micron) and green colors (4.5 \micron) in Fig.
\ref{irac-maps} reveal the stars, whereas the red color (8 \micron)
shows mainly the diffuse dust emission. Specifically, the origin of
diffuse emission at 3.6 \micron\ is likely to originate partly from
bound-free transitions in the ionized gas with some contribution from
the 3.29 \micron\ emission near molecular clouds (thought to be due to
C$-$H bond stretching in PAHs) and very small dust grain emission (e.g.
Engelbracht et al. 2006). The 4.5 \micron\ diffuse emission is a
combination of bound-free continuum and Brackett-$\alpha$ recombination
emission in {\sc H~ii} regions. Both the 5.8 and 8.0 \micron\ bands show
molecular material in the interstellar medium (ISM) and circumstellar
envelopes, as well as increasingly faint stellar photospheres.  The
origin of the 5.8 \micron\ diffuse emission is most likely very warm
dust emission ($T \sim 600$ K). It encompass the 5.6 and 6.2 \micron\
PAH band emission.  The origin of the 8 \micron\ diffuse emission is
most likely the very bright emission complex at 7 - 9 \micron\ thought
to be dominated by C$-$C stretching modes (with some contribution from
in-plane C$-$H bending) of the bonds in PAHs and very small carbonaceous
dust grains (J{\"a}ger et al. 2006).


\begin{figure}[t!]
\epsscale{1.05}
\plotone{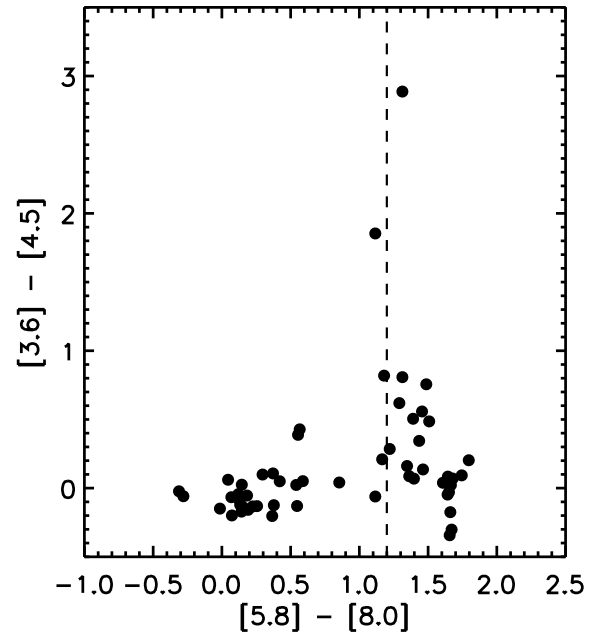} 
\caption{{\em Spitzer}/IRAC Color-color diagram of the sources detected   
in our photometry in the region of NGC~602/N~90. The vertical dashed   
line corresponds to the limit [3.6]$-$[8.0]$=$1.2 mag, set by Bolatto et
al. (2007) as a criterion for the selection of candidate YSOs similar to
those modeled by Whitney et al. (2003). We refine our selection of
candidate YSOs in the region to 22 point-sources, which are redder than
this limit.
\label{fig-ccd}}
\end{figure}

\begin{figure*}[t!]
\vspace*{2.5truecm}
\begin{center}
{\large The postscript file is removed due to size limitations. \\
To see the figure download the corresponding JPG file: ``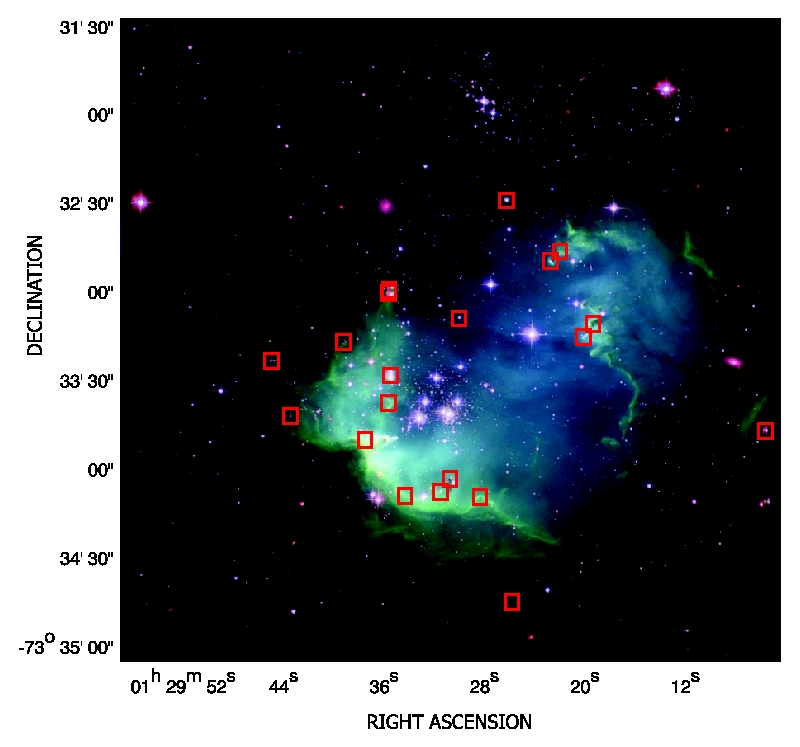''}
\end{center}
\vspace*{2.5truecm}
\caption{Color composite image of separate exposures made by ACS/WFC on 
{\em Hubble} Space Telescope in two broad-band filters, $F555W$ ($V$) and 
$F814W$ ($I$), and one narrow-band filter, $F658N$ (H\alp\ $+$ [N II]) of 
the region NGC~602/N~90. North is up and east to the left. The data of 
these observations are the subject of an investigation of the young 
stellar populations in the region (Schmalzl et al. 2007). Red boxes 
indicate the positions of the candidate YSOs detected with our IRAC 
photometry. This picture clearly indicates that current star formation 
takes place at the periphery of the central association NGC~602. Probably, 
star formation started at the association (where the highest concentration 
of PMS stars has been found; Schmalzl et al. 2007) and propagated 
outwards, mainly to the east and south, with events of ongoing clustered 
star formation still taking place along the dust ridges.\label{fig-ysomap}}
\end{figure*}

\section{Point-Sources Detected with {\em Spitzer} in NGC~602/N~90}

Our photometry revealed 57 objects detected in all four IRAC channels. 
44 sources were identified in channels 1, 2 and 3, but not in channel 4,
   while only two in channels 2,3 and 4 and not in channel 1. For the 
   classification of the sources we make use of the [3.6]$-$[8.0], [8.0] 
   color-magnitude diagram (CMD) in comparison with templates and models 
   compiled for a wide range of IR sources in the Milky Way. These 
   templates have been developed by Cohen (1993) for IRAS data on galactic 
   IR sources, adapted for the IRAC and 2MASS color classification scheme 
   for the Galactic Legacy Infrared Mid-Plane Survey Extraordinaire 
   (GLIMPSE) {\em Spitzer} project (Benjamin et al. 2003). We also make 
   use of the YSO models that have been developed by Whitney et al. (2004) 
   also for GLIMPSE. Both papers include several types of IR sources such 
   as main-sequence stars, red giants, O-rich and C-rich AGB stars, {\sc 
   H~ii} regions, and YSOs. In Fig. \ref{iractempl} ({\em left}), we show 
   the position of such sources in the [3.6]$-$[8.0], [8.0] CMD according 
   to the templates and models. We also show the corresponding CMD for the 
   point-sources we found in all four IRAC channels in the region of 
   NGC~602/N~90 (Fig. \ref{iractempl}, {\em right}). From the comparison 
   of the CMDs of Fig. \ref{iractempl} it is shown that there are 27 
   objects in our photometry with positions in the [3.6]$-$[8.0], [8.0] 
   CMD similar to galactic YSOs of class 0, I and II. However, Class 0 
   YSOs are very rare due to their short evolutionary phase, and such 
   objects would be extremely faint in the IRAC bands at the distance of 
   the SMC. Consequently, one should consider that, although few of our 
sources match the positions of Class 0 YSOs in the CMD according to the 
models, probably there are no YSOs of this class in our sample.

A criterion adopted by Bolatto et al. (2007) within the S$^3$MC Survey
(see also \S 1) for the selection of YSOs in their sample is based on
very red [5.8]$-$[8.0] color index ($>$~1.2 mag) of bright 5.8 \micron\
sources. Support for their selection is lent by the location of these
sources in the [3.6] $-$ [8.0], [8.0] CMD, and the color-color diagrams
in near-, mid- and far-infrared wavelengths as modeled by Whitney et al. 
(2003) for different objects. Bolatto et al. note that the highly
reddened 8.0 \micron\ sources occupy the region in those plots that is
predicted for late Class 0 to Class II YSOs with different geometrical
parameters.  We further select the most probable YSOs in the region of
NGC~602/N~90 by applying this criterion to the sources detected with our
photometry.  The [5.8]$-$[8.0], [3.6]$-$[4.5] color-color diagram of our
sources is shown in Fig. \ref{fig-ccd} with the limit of Bolatto et al.
(2007) for candidate YSOs overplotted as a vertical dashed line. There
are 24 sources in our photometry that have colors [5.8]$-$[8.0] $>$ 1.2
mag, meeting this selection criterion. All, except two of them, have
been selected as candidate YSOs also according to their positions in the
[3.6]$-$[8.0], [8.0] CMD (Fig. \ref{iractempl}). Consequently, we refine
the number of candidate YSOs in the region of NGC~602/N~90 to 22
sources, which meet both criteria.

The classification scheme of Bolatto et al. predicts only $\sim$ 2\% 
contamination of this part of the mid-infrared color-color diagram by B 
stars with 24 \micron\ emission (their Figure 12). Other types of evolved 
sources, such as carbon stars, RSGs and O-rich AGB stars, are located to 
bluer colors. In addition, Bolatto et al. estimated the contamination by 
background, unresolved galaxies, which can be seen throughout the SMC at 
infrared wavelengths. They expect 20 contaminants in their sample of 282 
candidate YSOs ($\simeq$ 7\%), which translates to 1.5 sources in our 
sample. However, detailed inspection of the candidate YSOs on the ACS 
image of NGC~602/N~90 revealed no suspicious background galaxies in our 
sample. The characteristics of the candidate YSOs revealed with our 
photometry in the region of NGC~602/N~90 are presented in Table 
\ref{tab-yso}, where we give the ID number (from our IRAC photometric 
catalog), the position and the magnitudes and corresponding photometric 
uncertainties in the four IRAC channels for each candidate YSO.

It should be noted that the classification of candidate YSOs cannot, in 
any case, be reduced to a single criterion, and therefore the technique  
described above should be considered a first-order selection. Moreover, 
taking into account that the selection of objects with 
[5.8]$-$[8.0]~$>$~1.2~mag may compromise the actual number of YSOs, this 
limit should be considered a tentative one. Indeed, Allen et al. (2004) 
plot known stars with YSO classifications in the IRAC color-color diagram 
(their Fig. 4) and their Class I YSOs clearly stretch to red [3.6]$-$[4.5] 
colors with [5.8]$-$[8.0] colors less than 1.2 mag. Our selection, based 
on Bolatto et al. (2007) scheme, apparently causes such sources to be left 
out, but its use is being considered in order to reduce as much as 
possible the number of misidentifications. Furthermore, sources with 
colors [3.6]$-$[4.5]~$\approx$~0.0~mag and [5.8]$-$[8.0]~$>$~1.2~mag are 
not well represented in Whitney et al. (2003, 2004) nor in Allen et al. 
(2004). Based on their colors, some of them are suspected to be small {\sc 
H~ii} regions, but there are no high resolution radio observations of N~90 
(unlike N~159 in the LMC; Jones et al. 2005), which would provide 
information on the ionizing flux for these sources and help us clarify 
their nature.

\begin{figure*}[t!]
\vspace*{2.5truecm}
\begin{center}
{\large The postscript file is removed due to size limitations. \\
To see the figure download the corresponding JPG file: ``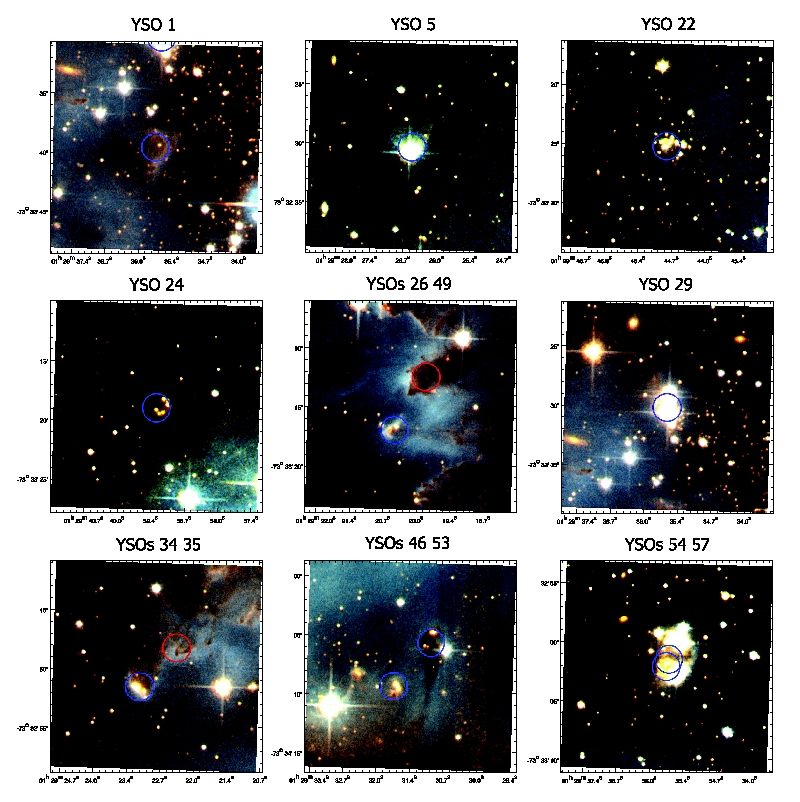''}
\end{center}
\vspace*{2.5truecm}
\caption{Color composite images constructed from our ACS observations of
regions $18\arcsec~\times~18\arcsec$ wide centered on the loci where   
candidate YSOs have been detected. Characteristic examples of 13 candidate
YSOs from Table \ref{tab-yso} have been selected. The drawn circles,
corresponding to radius 1\farcs2 indicate the positions of the candidate
YSOs. Blue circles denote candidate YSOs for which optical sources have
been found to coincide with, while red circles the ones for which no  
optical sources have been identified.\label{fig-ysosamples}}
\end{figure*}

\begin{deluxetable*}{cccccccccccccccc}

 
\tablecolumns{16}
\tablewidth{0pt}
\tabletypesize{\scriptsize}

\tablecaption{Candidate YSOs detected in all four IRAC channels in the
region of NGC~602/N~90. The numbering system of the YSOs is from our
photometric IRAC catalog. Celestial coordinates are given in J2000. The
magnitudes of each candidate YSO in all channels with the corresponding
errors are given in columns 4 through 11.  The last five columns refer to
the optical sources found with ACS to form possible compact clusters
within a radial distance of 1\farcs2 around each of the candidate YSOs.
The numbers of the detected optical sources, the magnitudes of the
brightest source (${\rm F555W}_{\rm bri}$, ${\rm F814W}_{\rm bri}$), as
well as the cumulative magnitudes of all stars around each candidate YSO
(${\rm F555W}_{\rm tot}$, ${\rm F814W}_{\rm tot}$) are given in columns
13, 14, 15 and 16 respectively. \label{tab-yso}}


\tablehead{
\colhead{ID} & 
\colhead{RA} & 
\colhead{DEC} &
\colhead{[3.6]} &
\colhead{$\pm$} &
\colhead{[4.5]} &
\colhead{$\pm$} &
\colhead{[5.8]} &
\colhead{$\pm$} &
\colhead{[8.0]} &
\colhead{$\pm$} &
\colhead{No} &
\colhead{${\rm F555W}_{\rm bri}$} &
\colhead{${\rm F814W}_{\rm bri}$} &
\colhead{${\rm F555W}_{\rm tot}$} &
\colhead{${\rm F814W}_{\rm tot}$} \\
\colhead{(1)} &
\colhead{(2)} &
\colhead{(3)} &
\colhead{(4)} &
\colhead{(5)} &
\colhead{(6)} &  
\colhead{(7)} &  
\colhead{(8)} &  
\colhead{(9)} &  
\colhead{(10)} & 
\colhead{(11)} & 
\colhead{(12)} & 
\colhead{(13)} & 
\colhead{(14)} &
\colhead{(15)} &
\colhead{(16)}
}

\startdata

1 &01:29:35.66&$-$73:33:39.61&15.90&0.22&16.24&0.30&13.64&0.24&11.99&0.15& 
2&24.62      &22.85  &24.46  &22.63  \\
2 &01:29:25.95&$-$73:34:46.69&16.52&0.23&16.82&0.32&14.17&0.31&12.50&0.19& 
5&25.45      &23.62  &24.66  &22.45  \\
5 &01:29:26.52&$-$73:32:30.31&16.27&0.15&16.45&0.21&14.57&0.29&12.91&0.15& 
2&17.84      &17.72  &17.82  &17.66  \\
22&01:29:44.77&$-$73:33:25.31&16.46&0.21&16.50&0.26&14.70&0.32&13.06&0.17&15&23.26      
&21.76  &21.82  &20.28  \\
24&01:29:39.23&$-$73:33:18.80&16.66&0.20&16.69&0.31&14.45&0.36&12.79&0.22& 
5&24.54      &21.85  &23.12  &20.93  \\
26&01:29:19.80&$-$73:33:12.46&16.06&0.23&16.04&0.29&13.84&0.30&12.17&0.17&  
&   &       &       &       \\
29&01:29:35.55&$-$73:33:30.20&14.13&0.09&14.09&0.11&12.22&0.13&10.61&0.08& 
8&17.49      &17.30  &16.73  &16.65  \\
34&01:29:23.08&$-$73:32:51.51&16.00&0.16&15.93&0.17&14.28&0.26&12.88&0.13& 
4&22.82      &20.95  &22.06  &20.13  \\
35&01:29:22.34&$-$73:32:48.17&16.67&0.27&16.60&0.36&14.63&0.35&12.95&0.22&  
&   &       &       &       \\
36&01:29:28.61&$-$73:34:12.14&16.88&0.31&16.80&0.42&15.32&0.50&13.68&0.29& 
1&25.15      &24.08  &       &       \\
38&01:29:34.08&$-$73:34:11.66&16.67&0.28&16.58&0.39&14.84&0.39&13.10&0.18&  
&   &       &       &       \\
42&01:28:47.20&$-$73:31:31.69&16.70&0.24&16.54&0.32&15.37&0.51&14.02&0.29&  
&   &       &       &       \\
43&01:29:41.38&$-$73:36:24.53&15.92&0.11&15.71&0.15&14.67&0.33&12.87&0.11&  
&   &       &       &       \\
45&01:29:43.34&$-$73:33:43.21&16.50&0.20&16.21&0.20&14.19&0.29&12.97&0.18& 
2&21.90      &21.22  &21.89  &21.16  \\
46&01:29:30.92&$-$73:34:05.74&16.51&0.23&16.17&0.27&14.80&0.38&13.36&0.21& 
1&22.45      &20.92  &       &       \\
49&01:29:20.44&$-$73:33:16.90&16.68&0.21&16.19&0.23&15.16&0.50&13.65&0.21& 
1&23.16      &21.55  &       &       \\
50&01:29:37.52&$-$73:33:51.34&15.00&0.18&14.50&0.20&13.09&0.22&11.70&0.19& 
1&20.96      &20.42  &       &       \\
51&01:29:06.45&$-$73:33:48.73&13.91&0.08&13.35&0.10&12.15&0.11&10.69&0.06&10&20.16      
&19.56  &20.05  &19.19  \\
52&01:29:30.21&$-$73:33:10.55&15.49&0.09&14.87&0.09&13.83&0.16&12.53&0.09& 
1&19.30      &18.62  &       &       \\
53&01:29:31.67&$-$73:34:09.39&15.36&0.14&14.60&0.13&13.42&0.17&11.94&0.07& 
1&26.44      &24.28  &       &\\
54&01:29:35.67&$-$73:33:02.11&13.07&0.07&12.26&0.05&11.18&0.08& 9.86&0.06& 
9&19.34      &18.54  &19.29  &18.42\\  
57&01:29:35.62&$-$73:33:01.52&15.14&0.20&12.26&0.05&11.18&0.08& 9.86&0.06& 
7&24.20      &22.37  &22.67  &20.78\\  

\enddata



\tablecomments{Objects No 54 and 57 being very close to each other, have
been identified as distinct sources only in the 3.6 \micron\ band. Our
photometry could not resolve them in IRAC channels 2,3 and 4, and
therefore they appear with the same corresponding magnitudes. Inspection
of their ACS image (Fig. \ref{fig-ysosamples}) shows indeed the existence
of multiple objects very close to each other. Around each of the
candidate YSOs 54 and 57 different PMS stars have been found with only
four common stars.  Objects No 42 and 43 are located outside the field
observed by ACS, and consequently it was not possible to identify any
optical sources for them.}


\end{deluxetable*}

\begin{figure}[]
\epsscale{1.25}
\plotone{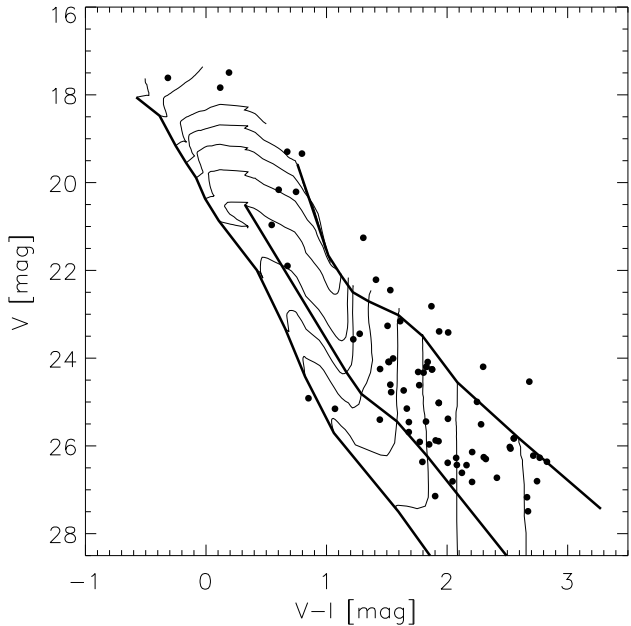}   
\caption{$V-I$, $V$ Color-Magnitude Diagram of all the optical sources
from our ACS photometry, the loci of which are found to coincide (within
a radial distance 1\farcs2) with the location of the candidate YSOs
identified with our IRAC photometry in the region of NGC~602/N~90.     
Evolutionary tracks for PMS stars from Palla \& Stahler (1999) for
masses 0.2, 0.4, 0.6, 0.8, 1.0, 1.2, 1.5, 2.0, 2.5, 3.0, 3.5, 4.0, 5.0
and 6.0 M\solar\ are overplotted. This diagram shows that these optical 
sources are PMS stars, which cover a wide range of masses. PMS isochrone
models for 0.1 and 5 Myr and the ZAMS are also plotted with thick solid
lines. These models limit the age of the PMS stars, and consequently the
timescale for the corresponding star formation events, to \lsim 5 Myr.
\label{fig-cmdpms}}
\end{figure}

\subsection{Optical Counterparts of the Candidate YSOs}

We use the results of the photometry by Schmalzl et al. (2007) on
ACS/WFC imaging of NGC~602 to identify the optical counterparts of the
candidate YSOs found with {\em Spitzer} located in this region. We
cross-correlate the positions of the candidate YSOs with the ones of all
stars found with ACS. For the comparison of the two catalogs we take
into account the fact that the resolution provided by IRAC is almost 25
times lower than the one of WFC, and therefore the selected area around
each candidate YSO to be searched for optical counterparts in most of
the cases revealed more than one optical sources. Specifically, we
selected a search circle with diameter of 2\farcs4, which corresponds to
the size of two IRAC pixels, a typical FWHM of the point-sources found
with our PSF photometry. In Table \ref{tab-yso} the results of this
search are given.  Specifically, in the last five columns of this table
we give the number of stars from the ACS photometry, found within a
radial distance of 1\farcs2 (2 IRAC pixels) around the center of each
candidate YSO. We also give the magnitude of the brightest optical
source in the filters $F555W$ ($V$) and $F814W$ ($I$) respectively, as
well as the corresponding total magnitudes from all optical sources
found within the searched area around each candidate YSO. In Table
\ref{tab-yso} there are five candidate YSOs for which no optical
counterpart was found. However, only three of them (No 26, 35 and 38)
are located within the ACS/WFC field-of-view for which the
cross-corelation was possible. The remaining two candidate YSOs (No 42
and 43) are located outside this field-of-view, since the selected area
observed by IRAC (Fig. \ref{irac-maps}) is a bit larger than the one
covered by ACS.

\section{Current Star Formation in the area of NGC~602/N~90}

\subsection{Clustered Star Formation Events}

As shown in Table \ref{tab-yso}, almost all of the detected candidate
YSOs in the area of NGC~602/N~90 are related to stars visible at optical
wavelengths. In half of the cases there are more than two optical
sources related to each candidate YSO (within a region corresponding to
two IRAC pixels). The loci of the detected candidate YSOs are shown
marked with red squares on a color composite image from the observations
taken with ACS/WFC of the area of NGC~602/N~90 in Fig. \ref{fig-ysomap}.
In this figure it is shown that there is a higher concentration of
candidate YSOs to the east and south, where also the nebula is brighter
and where the brighter 8\micron\ emission is found in the {\em Spitzer}
images (Fig.  \ref{irac-maps}). More detailed color composite images
from ACS imaging of the loci around 13 selected candidate YSOs are shown
in Fig.  \ref{fig-ysosamples}. Circles of radius corresponding to
1\farcs2 mark the centers of the candidate YSOs. In this figure it can
be seen that there are candidate YSOs coinciding with stellar
clusterings, which could be true compact clusters (objects No 22, 24,
29, 54/57). The nature of these compact clusters and if they are
signatures of on-going clustered star formation are two very important
issues.

We attempt to address these issues by plotting the CMD of all stars from 
our ACS photometry, that are associated with the candidate YSOs found with 
IRAC. The part of a $V-I$, $V$ CMD, which is expected to be occupied by 
pre-main sequence (PMS) stars of star-forming regions in the MCs observed 
with {\em Hubble} is fairly known (e.g. Gouliermis et al. 2006, Nota et 
al. 2006). This CMD for the optical sources related to the candidate YSOs 
in the region of NGC~602/N~90, plotted in Fig. \ref{fig-cmdpms}, shows 
that all of the stars found to coincide with the candidate YSOs in our 
sample are actually PMS stars, suggesting that either we probably observe 
the existence of compact PMS star clusters recently formed (or currently 
under formation) around typical YSOs, or that the sources identified as 
candidate YSOs {\em are} themselves unresolved young PMS clusters. In 
either case it is almost certain that these stellar concentrations 
demonstrate possible events of extragalactic clustered star formation on 
the act. If so, it would be interesting to know how long ago this 
formation took place. 

To answer this question, on the CMD of Fig. \ref{fig-cmdpms} we also plot 
PMS evolutionary tracks from the models by Palla \& Stahler (1999), and 
two indicative isochrones (for 0.1 and 5 Myr) and the Zero Age Main 
Sequence (ZAMS) from the same models. Simulations of the positions at the 
$V-I$, $V$ CMD of all PMS stars found in the region of NGC~602/N~90 by 
Schmalzl et al. (2007) indicate that reddening, variability and binarity 
are important factors, which cause a spread of the loci of PMS stars in 
this CMD and that even modest interstellar reddening (Schmalzl et al. find 
$E(B-V)\simeq$~0.04~-~0.06~mag for the association NGC~602) can make these 
stars appear younger (being ``pushed'' to redder colors). Considering 
these results and the plot of Fig. \ref{fig-cmdpms}, one may conclude that 
the PMS stars, which coincide with the candidate YSOs of our sample do not 
seem to be older than about 5 Myr, giving an upper limit to the time-scale 
within which these clusters were formed.

For a more detailed analysis one should concentrate on the stellar content 
of each of the individual clusters, but in most of the cases very low 
numbers of stars per cluster do not allow such an analysis. We selected 
the most populous among the identified clusters (the ones corresponding to 
candidate YSOs No 22, 29, 51 and 54) to plot their individual CMDs, but 
the stellar numbers were still not enough to define a more narrow age 
range. PMS stars are mainly red sources, and therefore near-infrared 
imaging with cameras equipped with Adaptive Optics (AO) systems would 
provide the necessary resolution for a richer sample of PMS stellar 
members in these clusters. Indeed, high resolution near-infrared 
observations of these clusters would certainly be very useful, because 
they would 1) provide better statistics with higher numbers of detected 
members, enough to plot the CMDs of individual clusters and 2) fill the 
gap between optical and mid-infrared wavelengths, providing near-infrared 
magnitudes for these PMS stars. The later information will offer the 
opportunity to construct accurate spectral energy distributions (SEDs) of 
candidate YSOs, to be compared to models.

As far as candidate YSOs that coincide with no more than two PMS stars 
concerns, in Fig. \ref{fig-ysosamples} we also show several different 
cases of ``single'' candidate YSOs (e.g. objects No 1, 5, 49, 53), but as 
their images show it is unclear if such sources are indeed single isolated 
objects or embedded clusters (as the faint magnitudes given in Table 
\ref{tab-yso} suggest), or even one massive embedded protostar. Modeling 
of the SEDs of such objects is quite difficult, in the sense that many 
different models seem to fit the SEDs, due to their small number of 
points, and near-infrared AO observations would certainly add three more 
points to the SEDs, throwing more light toward this direction as well. It 
is worth noting that Fig. \ref{fig-ysosamples} also demonstrates few 
interesting cases of candidate YSOs (objects No 26, 49, 34, 35, 46 and 
53), both with and without optical counterparts, which are located at the 
edge of features similar to the ``Pillars of Creation'' or ``Elephant 
Trunks'' in the galactic {\sc H~ii} region M~16 (the Eagle Nebula; Hester 
et al. 1996).

\subsection{Star-Formation Scenario for the Region of NGC~602/N~90}

The sample of candidate YSOs derived from our IRAC photometry in the 
region of NGC~602/N~90 (Table \ref{tab-yso}) is based on a single 
classification scheme. As a consequence, although it is almost certain 
that this sample does include true YSOs, there is always a possibility 
that it also includes sources misidentified as YSOs, like small {\sc H~ii} 
regions, or that it is quite incomplete not including all possible true 
YSOs, like the Class I YSOs in the sample of Allen et al. (2004). The 
identification of PMS stars in the areas of the candidate YSOs as their 
optical counterparts indicates that half of these objects are not single 
sources. But, considering the difficulty of modeling the SEDs of these 
objects, it is not straightforward to clarify which of them represent 
embedded proto-clusters. In the case of candidate YSOs coinciding with one 
or two PMS stars, there is no information if there should be more optical 
counterparts, which simply were not detected in our ACS photometry, or if 
these are indeed single proto-stellar objects.

In any case, the composite ACS image of the area of NGC~602/N~90 shown in 
Fig. \ref{fig-ysomap} provides evidence that sequential star formation 
does take place in the area and especially on the dust ridges of a cavity, 
which seems to extend outwards. This feature is most probably the product 
of photoionization by the young association, NGC~602, located almost at 
its center. The combination of both ACS (Fig. \ref{fig-ysomap}) and IRAC 
(Fig. \ref{irac-maps}) observations provides a more complete image of the 
star formation in the area. Specifically, taking into account the loci of 
the sources classified as candidate YSOs, the ISM features shown in the 
image from ACS, and the 8 \micron\ emission shown in the IRAC image, one 
may conclude that in the observed area the stellar association NGC~602 
ionizes the surrounding material giving birth to the {\sc H~ii} region, 
N~90, and producing a cavity in the remaining of its parental molecular 
cloud. This event triggers sequential star formation not later than 
$\sim$~5~Myr ago, which seems to propagate mostly to the east and south. 
This second star formation event may be characterized by higher a 
concentration of YSOs, most of which may be related to compact PMS 
clusters. This scenario is also supported by the study of the PMS 
population in NGC~602 and its surroundings by Schmalzl et al. (2007).

\section{Conclusions}

We present our results from {\em Spitzer}/IRAC photometry on the 
star-forming region NGC~602/N~90 in the Small Magellanic Cloud. 

We performed PSF photometry on IRAC images taken in four mid-infrared 
channels (3.6 \micron, 4.5 \micron, 5.8 \micron\ and 8.0 \micron) centered 
on the {\sc H~ii} region N~90. We identified 57 objects in all four 
channels and 44 in channels 1, 2 and 3. We classified the sources found in 
all four IRAC channels from their loci in the [3.6]$-$[8.0], [8.0] 
color-magnitude diagram, based on templates and models for different red 
sources in the Milky Way, and the [5.8]$-$[8.0], [3.6]$-$[4.5] color-color 
diagram, according to the classification scheme developed by Bolatto et 
al. (2007) for SMC YSOs.

According to our classification there are 22 candidate YSOs in the region 
of NGC~602/N~90. We use our results from PSF photometry in the filters $V$ 
and $I$ of the same region from imaging with the Wide-Field Channel of ACS 
on-board the {\em Hubble} Space Telescope (Schmalzl et al. 2007) to 
identify optical sources, which coincide with the candidate YSOs. All 
these optical sources are identified as PMS stars. We found that almost 
half of the candidate YSOs match with no more than two PMS stars, but 
examination of their loci on the ACS images does not exclude the 
possibility that there may be more highly embedded sources or not well 
resolved also in the optical. The remaining candidate YSOs coincide with 
more than 3 PMS stars, and four of them seem to include small compact PMS 
clusters of 7 (or more) stars. We found that these stars are probably 
formed earlier than 5 Myr ago, according to PMS evolutionary models. We 
could not find any optical counterpart for five YSOs, two of which are 
located outside the field observed with ACS.

From the positions of the detected candidate YSOs, we conclude that they
are probably the products of a sequential star formation process, which
presumably propagates mostly to the east and south, on the ridges of the
molecular cloud, as they formed from the photo-ionizing process of the
stellar association NGC~602 located in the center of the apparent cavity
of the cloud.

Finally, it should be mentioned that our sample of candidate YSOs 
may be contaminated by other objects like small {\sc H~ii} regions, or it 
may not include some true YSOs. Based on our results we stress the 
importance of high-resolution near-infrared imaging of the candidate YSOs 
in order 1) to resolve more faint red stellar sources and detect a higher 
number of related PMS stars, and 2) to fill the wavelength gap between 
optical and mid-infrared by providing near-infrared measurements of all 
detected sources, so that their SEDs will be thoroughly modeled, as well 
as of high-resolution radio observations, which will provide additional 
information on the ionizing flux for the identified objects.

\acknowledgments

Dimitrios A. Gouliermis kindly acknowledges the support of the German 
Research Foundation (Deutsche Forschungsgemeinschaft - DFG) through the 
individual grant 1659/1-1. Sascha P. Quanz kindly acknowledges financial 
support from the German Friedrich-Ebert-Stiftung. This paper is based on 
observations made with the Spitzer Space Telescope, which is operated by 
the Jet Propulsion Laboratory, California Institute of Technology under a 
contract with NASA, and on observations made with the NASA/ESA Hubble 
Space Telescope, obtained from the data archive at the Space Telescope 
Science Institute. STScI is operated by the Association of Universities 
for Research in Astronomy, Inc. under NASA contract NAS 5-26555. This 
research has made use of NASA's Astrophysics Data System, {\em Aladin} 
(Bonnarel et al. 2000), {\em WCSTools} (Mink 2001), and the SIMBAD 
database, operated at CDS, Strasbourg, France.




\begin{references}
\reference{} Allen, L.~E., et al.\ 2004, \apjs, 154, 363 
\reference{} Bakes, E., \& Tielens, A. G. G. M. 1994, ApJ, 427, 822
\reference{} Benjamin, R. A., et al. 2003, PASP, 115, 953
\reference{} Bolatto, A.~D., et al.\ 2007, ApJ, 655, 212 
\reference{} Bonnarel, F., et al.\ 2000, A\&AS, 143, 33 
\reference{} Clayton, G. C., Wolff, M. J., Sofia, U. J., Gordon, K. D., \&
Misselt, K. A. 2003, ApJ, 588, 871
\reference{} Cohen, M. 1993, AJ, 105, 1860
\reference{} Davies, R.~D., Elliott, K.~H., \& Meaburn, J.\ 1976, MmRAS,
81, 89
\reference{} Desert, F.-X., Boulanger, F., \& Puget, J. L. 1990, A\&A, 237, 215
\reference{} Engelbracht, C. W., et al. 2006, ApJ, 642, L127
\reference{} Fazio, G., et al. 2004, ApJS, 154, 10
\reference{} Gouliermis, D., Brandner, W., \& Henning, T.\ 2006, \apjl, 
636, L133 
\reference{} Hartmann, L., et al.\ 2005, ApJ, 629, 881 
\reference{} Henize, K.~G.\ 1956, ApJS, 2, 315
\reference{} Hester, J.~J., et al.\ 1996, \aj, 111, 2349 
\reference{} J{\"a}ger, C., Krasnokutski, S., Staicu, A., Huisken, F.,
Mutschke, H., Henning, T., Poppitz, W., \& Voicu, I.\ 2006, ApJS, 166,
557
\reference{} Jones, T.~J., Woodward, C.~E., Boyer, M.~L., Gehrz, R.~D., \& 
Polomski, E.\ 2005, \apj, 620, 731
\reference{} Hodge, P.\ 1985, PASP, 97, 530 
\reference{} Li, A., \& Draine, B. T. 2002, ApJ, 576, 762
\reference{} Massey, P.\ 1993, Massive Stars:  Their Lives in the 
Interstellar Medium, 35, 168 
\reference{} Massey, P.\ 2002, \apjs, 141, 81 
\reference{} Mink, D.~J.\ 2002, ASP Conf.~Ser.~281: Astronomical Data 
Analysis Software and Systems XI, 281, 169
\reference{} Palla, F., \& Stahler, S. W. 1999, ApJ, 525, 772
\reference{} Sauvage, M., Vigroux, L., \& Thuan, T. X. 1990, A\&A, 237, 296
\reference{} Schwering, P. B. W. 1989, A\&AS, 79, 105
\reference{} Simon, J. D., et al.\ 2007, submitted to ApJ
\reference{} Schmalzl, M. et al.\ 2007, to be submitted to ApJ
\reference{} Stanimirovi\'{c}, S., Staveley-Smith, L., van der Hulst, J.
M., Bontekoe, T. R., Kester, D. J. M., \& Jones, P. A. 2000, MNRAS, 315,
791
\reference{} Weingartner, J. C., \& Draine, B. T. 2001, ApJ, 548, 296
\reference{} Werner, M., et al. 2004, ApJS, 154, 1
\reference{} Whitney, B. A., Wood, K., Bjorkman, J. E., \& Cohen, M. 2003,
ApJ, 598, 1079
\reference{} Whitney, B., Indebetouw, R., Bjorkman, J. E., \& Wood, K.
2004, ApJ, 617, 1177
\reference{} Wolfire, M., Hollenbach, D., McKee, C. F., Tielens, A. G. G.
M., \& Bakes, E. L. O. 1995, ApJ, 443, 152
\end{references}
\end{document}